\documentclass[aip,rsi,reprint,graphicx]{revtex4-1} % for checking your page length
\usepackage{graphicx}
\usepackage{amsmath}
\makeatletter
\def\p@subsection     {\thesection {.}}
\makeatother

\begin{document}

%\preprint{}

\title{A High-Resolution Combined Scanning Laser- and Widefield Polarizing Microscope for Imaging at Temperatures from 4\,K to 300\,K} %Title of paper



\author{M. Lange}
%\email[matthias.lange@uni-tuebingen.de]
%\thanks{}
\author{S. Gu\'enon}
\author{F. Lever}
\author{R. Kleiner}
\author{D. Koelle}
\affiliation{Physikalisches Institut - Experimentalphysik II and Center for Quantum Science (CQ) in LISA$^+$, Universit\"at T\"ubingen, D-72076 T\"ubingen, Germany}


\date{\today}

\begin{abstract}
Polarized light microscopy, as a contrast-enhancing technique for optically anisotropic materials, is a method well suited for the investigation of a wide variety of effects in solid-state physics, as for example birefringence in crystals or the magneto-optical Kerr effect (MOKE). We present a microscopy setup that combines a widefield microscope and a confocal scanning laser microscope with polarization-sensitive detectors. By using a high numerical aperture objective, a spatial resolution of about 240\,nm at a wavelength of 405\,nm is achieved. The sample is mounted on a $^4$He continuous flow cryostat providing a temperature range between 4\,K and 300\,K, and electromagnets are used to apply magnetic fields of up to 800\,mT with variable in-plane orientation and 20\,mT with out-of-plane orientation. Typical applications of the polarizing microscope are the imaging of the in-plane and out-of-plane magnetization via the longitudinal and polar MOKE, imaging of magnetic flux structures in superconductors covered with a magneto-optical indicator film via Faraday effect or imaging of structural features, such as twin-walls in tetragonal SrTiO$_3$. The scanning laser microscope furthermore offers the possibility to gain local information on electric transport properties of a sample by detecting the beam-induced voltage change across a current-biased sample. This combination of magnetic, structural and electric imaging capabilities makes the microscope a viable tool for research in the fields of oxide electronics, spintronics, magnetism and superconductivity.  
\end{abstract}


\maketitle %\maketitle must follow title, authors, abstract and \pacs

% References should be done using the \cite and \label commands
\section{Introduction}
\label{sec:introduction}
The properties of ferroic materials and devices are strongly affected by their microscopic domain structure. Knowledge about the domains often plays a key role in the understanding and interpretation of integral measurements, which puts an emphasis on the importance of imaging techniques. Polarized light microscopy is an excellent tool for this purpose and has been successfully applied to ferromagnetic~\cite{McCord2015}, ferroelastic~\cite{Erlich2015} and ferroelectric~\cite{Ye2000,Tu2001} domain imaging. Alternative methods for imaging of magnetic domains include Bitter decoration~\cite{Bitter1932}, Lorentz microscopy~\cite{Jakubovics1997}, electron holography~\cite{Tonomura1983}, magnetic force microscopy~\cite{Hartmann1999}, scanning SQUID microscopy~\cite{Kirtley1999}, scanning Hall probe microscopy~\cite{Chang1992}, nitrogen vacancy center microscopy~\cite{Maertz2010}, X-ray magnetic circular dichroism~\cite{Schneider1997}, scanning electron microscopy (SEM)~\cite{Akamine2016}, and SEM with polarization analysis (SEMPA)~\cite{Scheinfein1990}. A comparison of most of these methods can be found in Ref.\onlinecite{Hubert1998}. Imaging of ferroelectric domains has also been accomplished by etching~\cite{Hooton1955}, nanoparticle decoration~\cite{Ke2007}, scanning electron microscopy~\cite{Aristov1984}, piezoresponse force microscopy~\cite{Güthner1992} and X-ray diffraction~\cite{Fogarty1996}. These techniques have been reviewed by Potnis~\textit{et al.}~\cite{Potnis2011} and by Soergel~\cite{Soergel2005}. 

Polarized light imaging provides a non-destructive, non-contact way to observe ferroic domains with sub-$\mu$m resolution and high sensitivity that can be carried out in high magnetic fields. The contrast for imaging of ferroelastic or ferroelectric domains arises from birefringence or bireflectance~\cite{Schmid1993, Grechishkin99}, which is a consequence of the anisotropic permittivity tensor of these materials.  Ferromagnetic domains, on the other hand, can be imaged via the magneto-optical Kerr effect~\cite{Kerr1877} (MOKE) or the Faraday effect~\cite{Faraday1846}. Both confocal laser scanning ~\cite{Webb96} and widefield~\cite{Inoue94} microscopy can be used for imaging with polarized light contrast. In confocal laser scanning microscopy the image is captured sequentially by scanning a focussed laser beam across the sample. A confocal pinhole eliminates light that does not originate from the focal volume. This results in a high depth discrimination, contrast enhancement and a 28\,\% increase in lateral resolution.  Widefield microscopy, on the other hand, has the advantage of faster acquisition rates and simultaneous image formation. 

The instrument discussed below is based on an earlier design by Gu\'{e}non~\cite{Guenon11} and combines a widefield- and a confocal laser scanning microscope with polarization-sensitive detectors. To study effects at low temperatures and in magnetic fields, the sample is mounted on a liquid-helium continuous flow cryostat offering a temperature range of 4\,K to 300\,K and magnetic fields up to 800\,mT with variable orientation can be applied. The confocal laser scanning microscope offers an additional imaging mechanism: a beam-induced voltage across a current-biased sample can be generated by the local perturbation of the laser beam. This beam-induced voltage can be used to extract local information on the electric transport properties of the sample~\cite{Kittel97,Li2015, Benseman2015,Sivakov2000,WangHB2009}. While several examples of low-temperature widefield~\cite{Kirchner73, Goa03, Golubchik2009} and laser scanning polarizing microscopes~\cite{Henn2013, Murakami2010, Matsuzaka2009} have been published, the instrument presented here stands out with regard to the versatility offered by combining widefield and confocal laser scanning imaging modes, the accessible temperature range, as well as the very high lateral resolution it provides at low temperatures.

This paper is organized as follows. Given the relevance for the design of the microscope, a brief overview of MOKE and Faraday effect is given in Section~\ref{sec:MOKE}. The cryostat and the generation of magnetic fields is described in Section~\ref{sec:Cryostat}. The widefield polarizing microscope is discussed in Section~\ref{sec:LTWPM} and a detailed description of the scanning laser microscope is presented in Section~\ref{sec:LTSPM}. Imaging of electric transport properties is addressed in Section~\ref{sec:electrictransport} and examples demonstrating the performance of the instrument are presented in Section~\ref{sec:Examples}.
\section{Magneto-optical Kerr effect (MOKE) and Faraday effect}
\label{sec:MOKE}
The observation of domains in magnetic materials relies mainly on two magneto-optical effects: the magneto-optical Kerr effect (MOKE) in reflection and the Faraday effect in transmission. Both effects lead to a rotation of the plane of polarization that depends linearly on magnetization and is caused by different refractive indices for left-handed and right-handed circularly polarized light (magnetic circular birefringence).

A distinction between three types of MOKE with regard to the orientation of the magnetization and the plane of incidence is made: polar, longitudinal and transverse MOKE, being sensitive to the out-of-plane magnetization component, the in-plane magnetization component along the plane of incidence and the in-plane magnetization component perpendicular to the plane of incidence, respectively. For linearly polarized light, both the longitudinal and the polar MOKE lead to a rotation of the plane of polarization upon reflection on the sample surface, while the transverse MOKE leads to a modulation of the reflected intensity. In addition to the rotation of the plane of polarization the longitudinal and polar MOKE also lead to elliptically polarized light caused by a difference in absorption for left- and right-handed circularly polarized light. Furthermore, the MOKE also depends on the angle of incidence (AOI). The polar MOKE is an even function of AOI and has the largest amplitude for normal incidence. The longitudinal MOKE is an odd function of AOI and increases with increasing AOI.

The Faraday effect can be observed when light is transmitted through transparent ferromagnetic or paramagnetic materials. It describes a rotation of the plane of polarization  that is proportional to the magnetization component along the propagation direction of the light and the length of the path on which the light interacts with the material. An important application of the Faraday effect are magneto-optical indicator films~\cite{Goernert2010} (MOIF), that can be used to image the stray field above a sample. These typically consist of a thin garnet film with in-plane anisotropy that is coated with a mirror on one side. The mirror side of the MOIF is placed in direct contact with the  sample under investigation and observed under perpendicular illumination with a polarizing microscope. The stray field of the sample leads to a deflection of the magnetization in the MOIF, that now has a component along the propagation direction of the light and thus becomes observable via the Faraday effect.

Additional magneto-optical effects that are rarely used for domain imaging are the Voigt effect and the Cotton-Mouton effect. 
A detailed description of the various magneto-optical effects can be found in Ref.~\onlinecite{Hubert1998, McCord2015}.

\section{Cryostat, Electromagnet}
\label{sec:Cryostat}
The cryostat and microscope are mounted on a vibrationally isolated optical table. Since the sample is fixed on a coldfinger, the microscope needs to be positioned relative to the sample with sub-$\mu$m resolution. A very sturdy, yet precise, positioning unit needs to be used. The microscope is connected to the cryostat via flexible bellows to allow for the positioning of the microscope. 
The cryostat is a liquid-helium continuous flow cryostat with the sample in vacuum. The temperature can be adjusted in a range of $T=4$\,K to $T=300$\,K. The coldfinger and sample holder have a diameter of $25.4$\,mm. Electrical contacts and mounting screws around the perimeter of the sample holder limit the available space for sample mounting. Samples with a dimension of up to $12\,\mathrm{mm}\times 12\,\mathrm{mm}$ can be conveniently mounted.  

Two electromagnets are used to generate in-plane magnetic fields of up to $B_\parallel=\pm800$\,mT and out-of-plane magnetic fields of up to $B_\perp=\pm20$\,mT.
The electromagnets are mounted on a frame that is separated from the rest of the setup to reduce the risk of vibrations being transferred to the microscope. The out-of-plane magnetic field is generated by a Helmholtz coil that achieves a field homogeneity of 0.2\,\% in a cylindrical volume of 5\,mm length in out-of-plane direction and 20\,mm diameter in the sample plane. The magnet, generating the in-plane magnetic field, can be rotated around the cryostat to allow for an adjustment of the orientation of the in-plane magnetic field. The in-plane magnetic field is homogeneous to within 1\,\% in a cubic volume with an edge length of $12\,\mathrm{mm}$. The magnetic fields for both magnets have been calibrated using a Hall probe with an accuracy of 2\,\%.
Due to spatial constraints limiting the size of the Helmholtz coil, the achievable out-of-plane magnetic field strength is limited to 20\,mT. The range of applications of the instrument could be enhanced by replacing the two electromagnets by a superconducting vector magnet that allows the application of magnetic fields with a strength $>1\,\mathrm{T}$ and variable orientation. 
\begin{figure*}
 \includegraphics{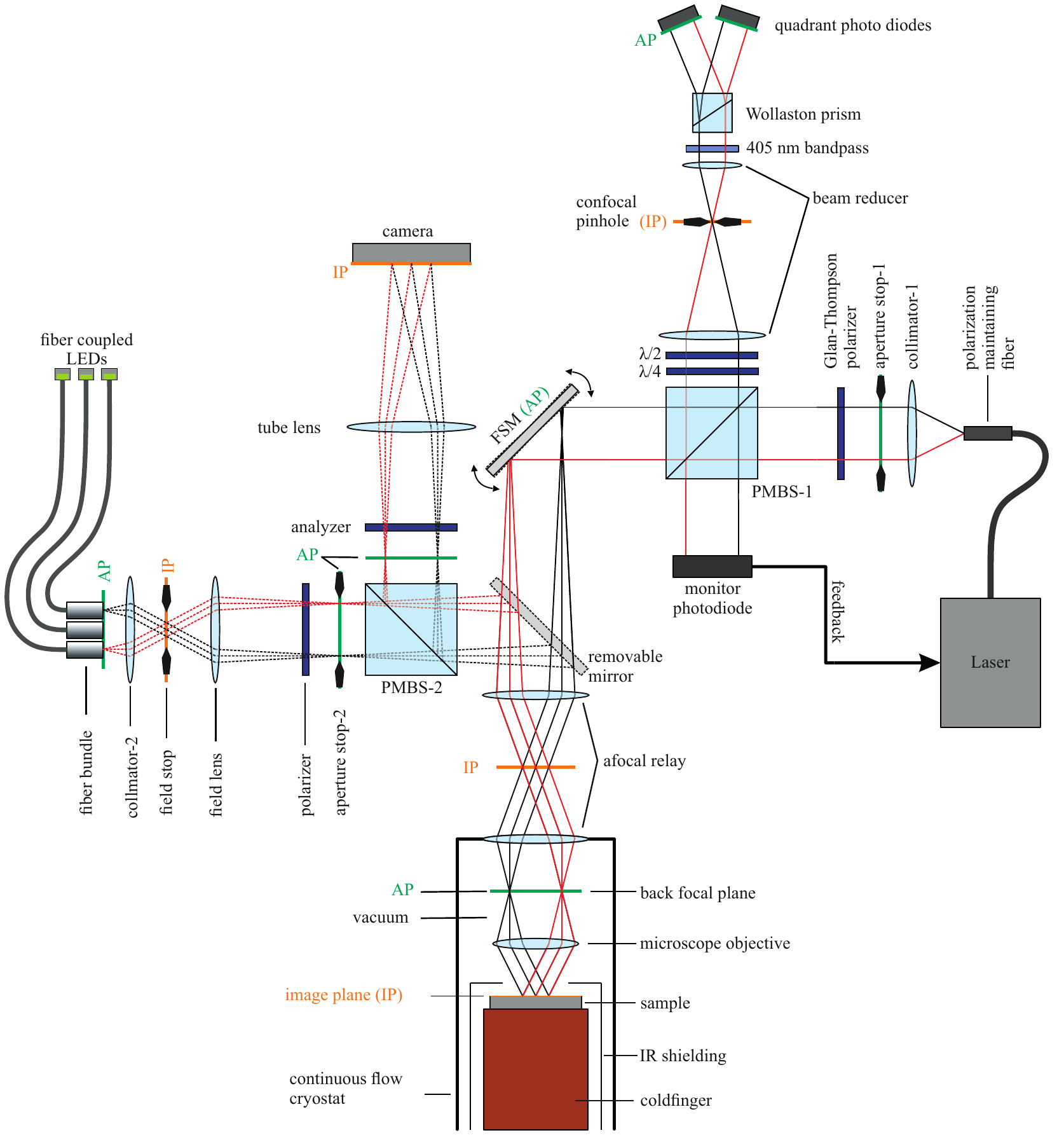}
 \caption{Schematic overview of the optical setup. Conjugate image planes (IP) are displayed in orange, conjugate aperture planes (AP) in green. The setup combines two imaging paths that can be selected via the removable mirror. A widefield polarizing microscope (mirror inserted), in which the sample is imaged onto the sensor of the camera, and a scanning polarizing microscope (mirror removed), that uses a fast-steering mirror (FSM) to scan a laser-beam across the sample. \label{fig:ttrpm}}
\end{figure*}
\section{Low-Temperature Widefield Polarizing Microscope (LTWPM)}
\label{sec:LTWPM}
The optical setup combines two imaging paths, a scanning polarizing microscope and a widefield polarizing microscope; the latter can be selected by inserting a mirror into the optical path. A schematic drawing of the imaging setup is displayed in Fig.~\ref{fig:ttrpm}. Conjugate planes to the image plane are denoted as image planes (IP) and shown in orange. Conjugate planes to the back focal plane (BFP) of the microscope objective are denoted as aperture planes (AP) and shown in green. AP and IP have a reciprocal relationship~\cite{Inoue94}: rays that are parallel in one set of planes are focussed in the other set of planes. The position of a point in the IP translates to an angle in the AP and the angle of a ray in the IP corresponds to a point in the AP.

In this Section we begin by giving a general description of the optical setup of the low-temperature widefield polarizing microscope (LTWPM) before we proceed with a detailed description of the components and their function.
The LTWPM can be used by inserting the removable mirror, as indicated by the broken lines in the ray diagram (Fig.~\ref{fig:ttrpm}). The illumination follows a Koehler scheme~\cite{Koehler1893} and the sample is illuminated through the microscope objective. The light source for the LTWPM is realized by fiber-coupled light-emitting diodes (LED), that are combined into a fiber bundle. The fiber bundle end face is imaged into the back focal plane of the microscope objective. To achieve this, the fiber output is collimated (collimator-2) and the field lens is used to image the fiber ends into an AP at the position of an interchangeable aperture stop (aperture stop-2). A rotatable polarizer in front of the aperture stop is used to define the plane of polarization. A field stop, aligned with the shared focal plane of collimator-2 and the field lens, can be used to confine the illuminated sample area. After passing a polarization maintaining beam splitter (PMBS-2), the light is reflected by the removable mirror onto the part of the setup that is shared with the scanning polarizing microscope. The light passes an afocal relay, which is used to extend the optical path into the cryostat and to image the light source into the BFP of the microscope objective. The afocal relay is discussed in detail in Section~\ref{subsec:relay}. After passing the afocal relay, the light is focussed onto the sample by the microscope objective. The light is reflected back from the sample, passes the microscope objective, afocal relay and removable mirror and is deflected by the PMBS-2. Subsequently, it passes the rotatable analyzer, that is used to adjust the polarized light contrast. The tube lens forms the image on the sensor of the low noise, high dynamic range sCMOS camera. 
\subsection{Light Source}
\label{subsec:illumination}
\begin{figure}
 \includegraphics{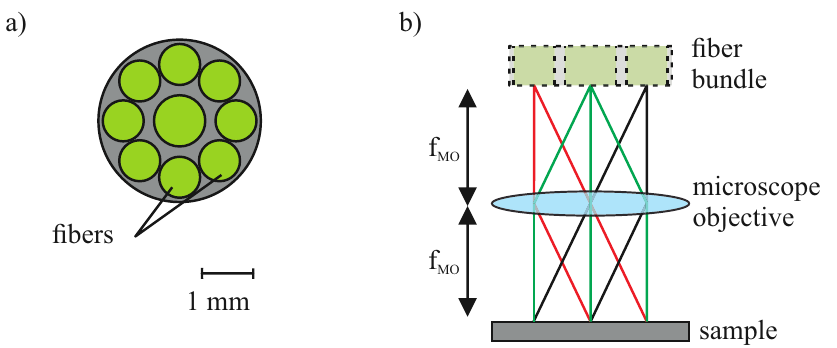}
 \caption{a) Cross-section of the fiber-bundle: The central fiber is aligned with the optical axis. The eight surrounding fibers are equally distributed on a concentric circle. b) The fiber end faces are imaged into the microscope objective back focal plane. The light originating from each fiber reaches the sample at a specific angle. The angle of incidence for the illumination can be selected by controlling the LEDs output power individually. \label{fig:fiberbundle} }
\end{figure}

The light source consists of nine high-power LEDs with a dominant wavelength of 528\,nm that are coupled into multimode fibers. The LEDs are temperature-stabilized by thermo-electric coolers and driven with highly stable current sources. The use of LEDs offers the advantage of low noise, compact size and excellent stability. The nine fibers are combined into a fiber bundle consisting of one central fiber of 1\,mm diameter and eight surrounding fibers of 0.8\,mm diameter as shown in Fig.~\ref{fig:fiberbundle}. A similar illumination concept has been developed by Soldatov \textit{et al.}~\cite{Soldatov2017,Soldatov2017b}. Since the LEDs can be controlled individually and the fiber ends are imaged into the back focal plane of the microscope objective, it is possible to switch between different angles of incidence. In magneto-optical imaging, the plane of incidence together with the plane of polarization, is used to adjust the sensitivity of the instrument to the polar, longitudinal or transverse MOKE~\cite{McCord2015,Stupakiewicz2014}. As was demonstrated by Soldatov \textit{et al.}, this illumination concept can be used to separate longitudinal and polar MOKE and to achieve a contrast enhancement. The LEDs can be pulsed with a frequency of more than 1\,MHz, which enables time-resolved imaging with $\mu$s temporal resolution.
\subsection{Microscope Objective}
\label{subsec:microscopeobjective}
In order to achieve a high resolution it is necessary to use a microscope objective with a high numerical aperture (NA). These typically have short working distances, which do not allow for the use of a vacuum window in front of the objective. Also, a window in front of the objective would introduce unwanted variations to the polarization of the light. Therefore, the microscope objective needs to be mounted inside the cryostat. This imposes the requirement for the microscope objective to be vacuum-compatible. At the same time, the possibly large difference in temperature between the microscope objective and the sample prohibits the use of immersion objectives or objectives with excessively short working distances, which limits the choice of high NA objectives. We decided on a commercial infinity corrected microscope objective with a NA of 0.8 and a focal length of $f_{MO}=4\,\mathrm{mm}$. It has a working distance of $WD=1\,\mathrm{mm}$ and is designed for polarized light microscopy with lenses that have been mounted strain-free.
It has a field of view (FOV) of 500\,$\mu$m diameter and features high transmission of about 80\,\% at 405\,nm and 87\,\% at 528\,nm wavelength. The exit pupil diameter is 6.4\,mm. Because it is also used for the scanning polarizing microscope, it needs to be suitable for confocal scanning.
\subsection{Camera, Polarization Sensitivity and Resolution}
\label{subsec:magnification}
The small variations in polarization, caused for example by magneto-optical effects, only lead to a very weak modulation of the intensity reaching the camera. Therefore, cameras with low noise and high dynamic range are required for imaging in polarized light microscopy. The camera we use is a thermoelectrically cooled sCMOS camera with 2048\,x\,2048 pixels, a pixel size of $6.5\,\mu$m\,x\,$6.5\,\mu$m and a high quantum efficiency of $QE \approx 80\,\%$ at a wavelength of $\lambda=528$\,nm. The full-well capacity (maximum number of photo-electrons) that can be stored on a pixel is $FW=30000$. The camera's noise specifications are given as a number of electrons. It features low median read noise of $N_r=0.9$ and a dark current of $D=0.10\,\mathrm{s}^{-1}$ per pixel. The dynamic range of 33000:1 is sampled with a bit depth of 16\,bit.

The achievable polarized light contrast depends on the extinction ratio of the polarizer and analyzer, as well as on the depolarizing effects occuring at the optics in between the polarizer and analyzer. 
Due to their compact size and moderate cost, we use polarizers based on oriented silver nanoparticles featuring an extinction ratio exceeding $\kappa = 1\cdot10^{-5}$ at a wavelength of 528\,nm. Depolarizing effects occuring at the optical elements between polarizer and analyzer~\cite{Shribak2002} lead to a reduction of the extinction ratio to a value around $\kappa=1\times10^{-2}$. 

We estimated the polarization sensitivity based on the extinction ratio of the microscope and the camera specifications. The signal is given by the number of photo-electrons $N=M \cdot B \cdot P\cdot t \cdot QE$ that are generated by a photon flux $P$ during the exposure time $t$ on a pixel with the quantum efficiency $QE$ for integrating over $M$ exposures and binning of $B$ pixels. The noise sources contributing to the overall noise are: The signal related shot noise $\delta_{shot}=\sqrt{N}$, dark noise $\delta_{dark}=\sqrt{M\cdot B \cdot D\cdot t}$ due to thermally generated electrons and readout noise $\delta_{read}=\sqrt{M \cdot B} \cdot N_r$. By adding the noise contributions in quadrature we derive the signal-to-noise ratio
\begin{equation}
	\mathrm{SNR} = \sqrt{M \cdot B} \cdot \frac{P\cdot t \cdot QE}{\sqrt{P\cdot t \cdot QE +D\cdot t+N_r^2}}~.
\end{equation}
\begin{figure}
 \includegraphics{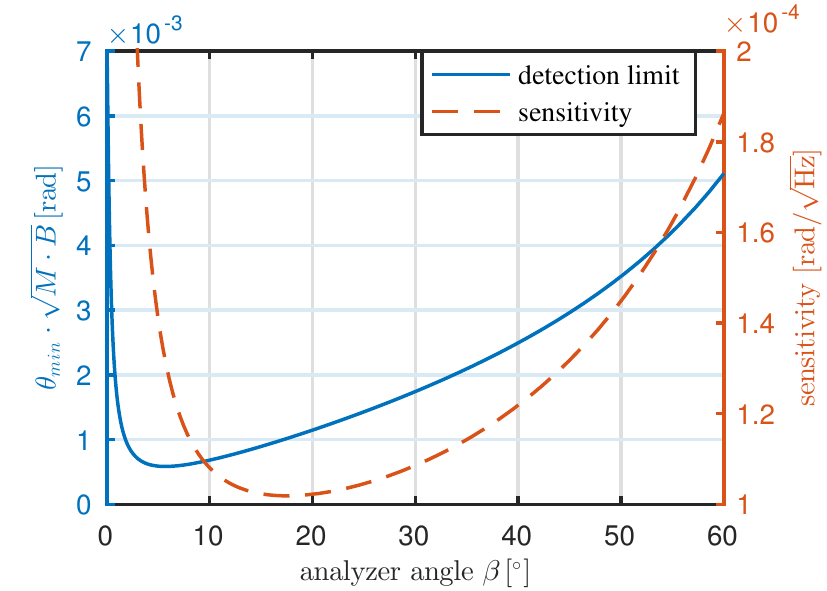}
 \caption{Detection limit $\theta_{min}\sqrt{M \cdot B}$ (solid blue line) and sensitivity (dashed red line) as a function of analyzer angle. \label{fig:sens_vs_M}}
\end{figure}
The SNR increases with the square root of the number of integrated images and binned pixels.
In polarized light imaging, the signal $N$ has to be substituted by the difference in photo-electrons that is generated by a rotation of the plane of polarization by an angle $\theta$. This is a function of the analyzer angle $\beta$ and is given by $N_{pol}= N \cdot (1-\kappa)(\mathrm{sin}^2(\beta + \theta)-\mathrm{sin}^2(\beta))$. The shot noise has to be replaced by the expression $\delta_{shot}=\sqrt{N \cdot ((1-\kappa) \cdot \mathrm{sin}^2(\beta + \theta) + \kappa)}$. If we further consider that the full well capacity $FW$ (maximum number of photo-electrons) of a camera pixel is limited we see that the highest SNR for a single exposure is achieved after an exposure time of $t_1=FW/(P \cdot QE ((1-\kappa)\mathrm{sin}^2(\beta + \theta)+\kappa))$. Now we can evaluate the smallest angle $\theta_{min}$, the detection limit, that can be measured with a SNR of 1 as a function of analyzer angle $\beta$ for a exposure time of $t_1$. As can be seen in Fig.~\ref{fig:sens_vs_M}, a detection limit of $\theta_{min}\sqrt{M \cdot B}= 6\cdot 10^{-4} \,\mathrm{rad }$ is reached for an analyzer angle of $\beta=5.7\,^{\circ}$. However, since the exposure time increases with decreasing analyzer angle $\beta$, the highest sensitivity of $\theta_{min}\cdot\sqrt{t}=1.0\cdot 10^{-4} \,\mathrm{rad/\sqrt{Hz}}$ is realized for an analyzer angle of $\beta=17.6\,^{\circ}$. 

The magnification $m_{wf}$ of the image, formed by the tube lens on the camera sensor, is determined by the ratio of the 200\,mm focal length $f_{TL}$ of the tube lens and the 4\,mm focal length $f_{MO}$ of the microscope objective, divided by the angular magnification $M_{AFR}$ of the afocal relay (see Sec.~\ref{subsec:relay}). This results in a total  magnification of $m_{wf}=f_{TL}/(f_{MO}M_{AFR})=27$.  
The diffraction limited resolution at a wavelength of 528\,nm and for the NA of 0.8 is given by the Rayleigh criterion $d_{min}=0.61\cdot\lambda/NA=403\,\mathrm{nm}$, which equals to about $11\,\mu$m on the camera sensor. The image is sampled with a resolution of $d_{px}=6.5\,\mu$m on the image sensor. Therefore, the achievable resolution is effectively limited by the Nyquist sampling theorem and is given by 
\begin{equation}
	d_{eff}=\frac{2\cdot d_{px}}{m_{wf}}=481\,\mathrm{nm}~.
\end{equation}
\section{Low-Temperature Scanning Polarizing Microscope (LTSPM)}
\label{sec:LTSPM}
In principle, two different scanning mechanisms can be utilized in scanning laser microscopy. First, there is mechanical scanning, where either the microscope or the sample is moved with piezo drives or stepper motors.
The second possibility is optomechanical scanning, where the optical path is modified by a movable component, for example a mirror, so that the spot changes its position on the sample.
We decided on the use of optomechanical scanning, because it offers several advantages over mechanical scanning. While providing a very good spatial resolution, optomechanical scanning has the advantage of a larger field of view and enables faster acquisition rates. Another benefit is that electromagnetic noise, that is generated by piezo drives and could disturb measurements of the electric transport properties, is avoided. Additionally, if the sample is to be mounted in a cryostat, good thermal coupling of the sample to the coldfinger is easily achieved.
However, in optomechanical scanning, greatest care has to be taken to ensure that the modification of the lightpath doesn't cause abberrations that prevent diffraction limited imaging or lead to a non-negligible influence on the polarization state. Therefore, all the components that are used in the beam scanning part of the microscope need to be polarization maintaining and designed for confocal scanning.

Hereafter, a  general description of the LTSPM setup, shown in Fig.~\ref{fig:ttrpm}, is given and subsequently, the individual components are discussed in detail.
The light source for the LTSPM is a 405\,nm wavelength laser diode coupled into a single-mode polarization maintaining fiber. The fiber output is collimated and the beam diameter is defined by an adjustable aperture stop. The plane of polarization is defined by a Glan-Thompson polarizer. The light passes a polarization-maintaining beamsplitter (PMBS-1) with a splitting ratio of 50:50, the light that is deflected by the beamsplitter is captured by a photodiode and is used as feedback for the stabilization of the laser intensity. The transmitted light is deflected by a two-axis fast steering mirror (FSM), which is used to set the position of the laser spot on the sample. The mirror is in a conjugate plane to the backfocal plane (aperture plane AP) of the infinity corrected microscope objective. An afocal relay is used to image the mirror onto the backfocal plane. The angular position of the FSM thereby defines the XY-position of the spot on the sample. This results in a telecentric illumination of the sample, where the beam is reflected and consequently passes the microscope objective, afocal relay and FSM in opposite direction. The light is then deflected by the PMBS-1. A quarter-wave plate ($\lambda$/4) is used to correct for the ellipticity of the reflected beam's polarization and a half-wave plate ($\lambda$/2) can be used to rotate the plane of polarization. A beamreducer is necessary to match the beam diameter to the size of the photodiodes. In the intermediate image plane of the beamreducer, a pinhole aperture is inserted to make the microscope confocal. The beam then passes a 405\,nm center wavelength band-pass filter and is split into two perpendicularly polarized beams by the Wollaston prism. These two beams are detected using two quadrant photodiodes. 
\subsection{Light Source}
\label{subsec:lightsource}
The key requirements for the light source are long-term stability and low noise. The light source consists of a temperature-stabilized diode laser with a wavelength of $\lambda = 405$\,nm and a maximum output power of $P_{max} = 50$\,mW, which is coupled into a polarization maintaining single-mode fiber. The control electronics of the laser diode operate on a battery power supply to reduce noise and the output power is controlled using the photodiode at the PMBS-1 as feedback. The laser power can be modulated with frequencies up to $f_{max} = 1$\,MHz.
\subsection{Fast Steering Mirror (FSM)}
\label{subsec:FSM}
Laser-beam scanning can be accomplished by means of galvanometric scanners~\cite{Montagu2011}, acousto-optical deflectors~\cite{Lv2006} or fast steering mirrors. Galvanometric scanners and acousto-optic deflectors provide angular displacement of the beam about a single axis. Therefore, it is necessary to use two separate scanners in a perpendicular orientation to achieve XY-scanning. Unless additional relay optics are used in between the two scanners, this results in linear displacement of the laser-beam from the optical axis. The main advantage of fast steering mirrors is that they provide angular displacement about two perpendicular axes in a single device. Thus, the mirror can be placed in a conjugate plane to the backfocal plane of the microscope objective. This results in pure angular displacement of the beam and a telecentric illumination of the sample.
We use a two-axis fast steering mirror with a mechanical scan range of $\pm{1.5}\,^\circ$ and an angular resolution of better than $2\,\mu$rad.

As has been discussed by Ping~\textit{et al.}~\cite{Ping95}, reflection of a linearly polarized laser beam at a mirror surface will lead to depolarization, caused by the difference in reflectivity and phase for s- and p-polarized light, if the beam is not purely s- or p-polarized. Pure s- or p-polarization can only be realized for a single scan axis. Since we use a two-axis scan mirror, depolarizing effects cannot be avoided. To minimize these effects, we use a dielectric mirror with a phase-difference of less than $5\,^\circ$ and a reflectivity-difference below $0.05\,\%$ for s- and p-polarized light at angles of incidence of $45 \pm 3^\circ$ and a wavelength of 405\,nm.
\subsection{The Afocal Relay (AFR)}
\label{subsec:relay}
In confocal laser beam scanning microscopy, the laser spot is scanned across the sample by pivoting the laser beam in the back focal plane of the microscope objective. To achieve this, the scan mirror is imaged into the backfocal plane using an afocal relay. In our case, the microscope objective is mounted inside the cryostat and consequently a vacuum window is needed at some point before the beam enters the microscope objective. We decided to use one of the lenses of the afocal relay as vacuum window.

An afocal relay is realized by combining two focal systems in such a way that the rear focal point of the first focal system is coincident with the front focal point of the second focal system. A collimated beam entering the first focal system will exit the second focal system as a collimated beam. The linear magnification $m=f_2/f_1$ and the angular magnification $M=f_1/f_2$ are determined by the equivalent focal lengths $f_1$ and $f_2$ of the first and second focal system, respectively. Cemented achromatic doublets are often used to realize relay lenses. However, they are not well suited for confocal imaging, mainly because of the astigmatism and field curvature they introduce~\cite{Ribes2000, Negrean2014}. Therefore, more complex lens systems need to be used, which are designed to correct these aberrations. We decided on air spaced triplets as a starting point for the two lens groups that make up the afocal relay, since they offer sufficient degrees of freedom to be made anastigmatic~\cite{Laikin2006}. However, a fourth lens, which acts as the cryostat window, has to be added to the lens group facing the microscope objective, because the mechanical stress exerted by the vacuum needs to be handled.

\begin{figure*}
 \includegraphics{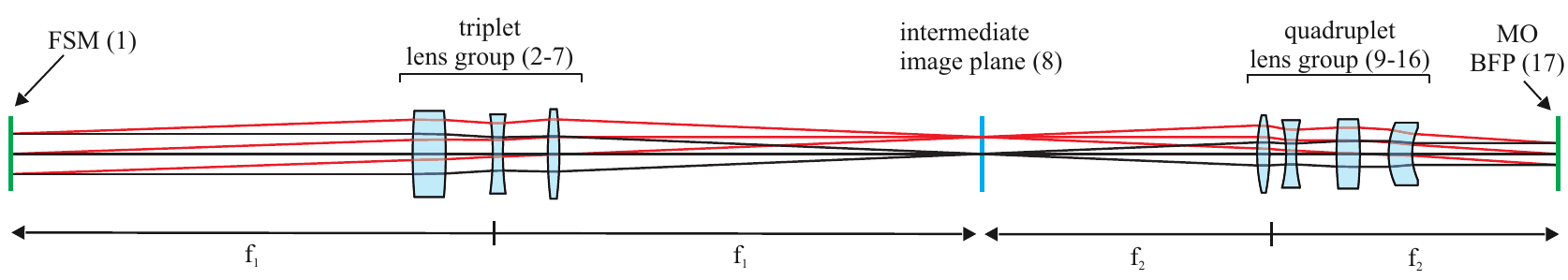}
 \caption{Layout of the afocal relay. Two lens groups, one triplet and one quadruplet, are used to relay the fast-steering mirror (FSM) onto the microscope objective's back focal plane (MO-BFP). The FSM is placed in the entrance pupil on the left side, the MO-BFP is coincident with the exit pupil on the right side. The rightmost lens of the quadruplet, a meniscus lens, is used as the cryostat window. The surface numbering is consistent with table~\ref{tab:relay} \label{fig:relay}}
\end{figure*}

The afocal relay has been specifically designed for use with the microscope objective and scan mirror.
It has an angular magnification of $M_{AFR}=1.85$, so that the range of the scan mirror is matched to the field of view of the microscope objective, and the rear aperture of the microscope objective will be completely filled using a laser beam with a diameter of 12\,mm. 
The relay consists of two lens groups: A triplet of two bi-convex and one bi-concave lenses and a quadruplet of two bi-convex, one bi-concave and one meniscus lens (Fig.~\ref{fig:relay}). The two lens groups are aligned as a 4f system, with the scan mirror placed in the front focal plane of the triplet lens group and the microscope objective back focal plane being coincident with the back focal plane of the quadruplet lens group. The lens parameters for the afocal relay are given in table~\ref{tab:relay}.

\begin{table}
 \caption{Lens data for the AFR: Radius of curvature of the surfaces defining the optical system and their separation along the optical axis. Material specifies the medium that fills the space between the current surface and the next surface. The surface numbering is consistent with Fig.~\ref{fig:relay}.  \label{tab:relay} }
 \begin{tabular}{| p{1cm} | p{1.4cm} | p{1.4cm} | p{1.7cm} | p{2cm}|}
 \hline
 surface no. & radius [mm] & separation [mm] & material & description \\ \hline
 1 & inf & 119.637 & air & entrance pupil\\ \hline
 2 & 159.604 & 9.809 & N-LASF44 & bi-convex\\ \cline{1-4}
 3 & -159.604 & 14.000 & air & lens\\ \hline
 4 & -89.906 & 3.000 & SF1 & bi-concave\\ \cline{1-4}
 5 & 89.906 & 13.356 & air & lens\\ \hline
 6 & 113.967 & 3.431 & N-BAF10 & bi-convex \\ \cline{1-4}
 7 & -113.967 & 125.237 & air & lens\\ \hline
 8 & inf & 82.248 & air & intermediate image plane \\ \hline
 9 & 50.810 & 3.647 & N-LAF34 & bi-convex\\ \cline{1-4}
 10 & -61.500 & 4.945 & air & lens\\ \hline
 11 & -43.268 & 3.000 & SF1 & bi-concave\\ \cline{1-4}
 12 & 43.268 & 12.000 & air & lens\\ \hline
 13 & 104.606 & 6.928 & N-LAK33A & bi-convex\\ \cline{1-4}
 14 & -104.606 & 8.549 & air & lens \\ \hline
 15 & 22.402 & 7.000 & SF57HHT & meniscus lens and cryostat\\ \cline{1-4}
 16 & 18.607 & 42.571 & vacuum & window\\ \hline
 17 & inf & & & exit pupil\\ \hline
 \end{tabular}
\end{table}
The AFR design has been optimized using a ray tracing software, Zemax OpticStudio\cite{Zemax}, to reduce all aberrations, especially astigmatism, coma, spherical aberration, field curvature and distortion for wavelengths of $\lambda=405$\,nm and $\lambda=528$\,nm. Since the relay needs to be polarization maintaining, the design has also been optimized in this regard, including strain-free mounting of the lenses. The meniscus lens acts as the cryostat window and is, unavoidably, under considerable mechanical stress. This lens was not only optimized with regard to its optical performance but also with regard to the mechanical demands imposed by the vaccuum. This lens has been fabricated from SF57HHT glass, which has an extremely low stress-optical coefficient, to minimize stress birefringence. Optical performance of the relay is diffraction limited over the entire scan range, which is essential for confocal imaging.
\subsection{Beam Reducer, Confocal Pinhole and Resolution}
\label{subsec:beamreducer}
The beam reducer is built from two commercial achromatic doublets with a design wavelength of 405\,nm. They have equivalent focal lengths of $f_{BR1}=125\,\mathrm{mm}$ and $f_{BR2}=25\,\mathrm{mm}$, so that the exiting beam is matched to the photodiode diameter of 2.5\,mm. The confocal pinhole aperture is mounted in the intermediate image plane of the beamreducer, which is a conjugate plane to the sample, and consequently blocks light that is not originating from the focal volume. The pinhole diameter $d_{ph}$ is determined by the diameter of the airy disc and the magnification $m_{cf}=f_{BR1}/(f_{MO}\cdot M_{AFR})$ of the microscope. It is given in Airy units ($\mathrm{AU}$), with $1\,\mathrm{AU}$ being the diameter of the image of the airy disc in the intermediate image plane of the beamreducer. The 125\,mm achromat was selected, so that $1\,\mathrm{AU}=10\,\mu\mathrm{m}$, with 
\begin{equation}
	1 \mathrm{AU}=\frac{1.22\cdot\lambda}{NA}\cdot m_{cf}~.
\end{equation}	

The resolution in confocal microscopy is increased by 28\,\% in comparison to widefield microscopy. In widefield microscopy, the resolution is given by the distance between two points for which their point spread functions (PSF) can be distinguished and is expressed by the Rayleigh criterion $d_{min}=0.61\cdot\lambda/NA$. In the Rayleigh criterion, the point-spread function is described by the Airy disk. Two point sources are considered to be resolvable, if the first minimum of the Airy disk of one point coincides with the global maximum of the other. In this case, the combined intensity profile shows a dip of $\approx26\,\%$ between the maxima corresponding to the two points. The increase in resolution in confocal microscopy originates from the fact, that the confocal volume is defined by the product of the illumination PSF and the convolution of the detection PSF with the pinhole~\cite{Webb96}. For a pinhole with a diameter of $0.5\,\mathrm{AU}$, this results in a function with a sharper peak compared to the widefield PSF. In this case, the confocal resolution $d_{cf}$ is given by 
\begin{equation}\label{eq:cfres}
	d_{cf}=\frac{0.44\cdot\lambda}{NA}=\frac{0.44\cdot405\,\mathrm{nm}}{0.8}=222\,\mathrm{nm}~.
\end{equation}
Using pinholes smaller than $0.5\,\mathrm{AU}$ does not increase the resolution, but deteriorates the SNR.

\begin{figure}
 \includegraphics{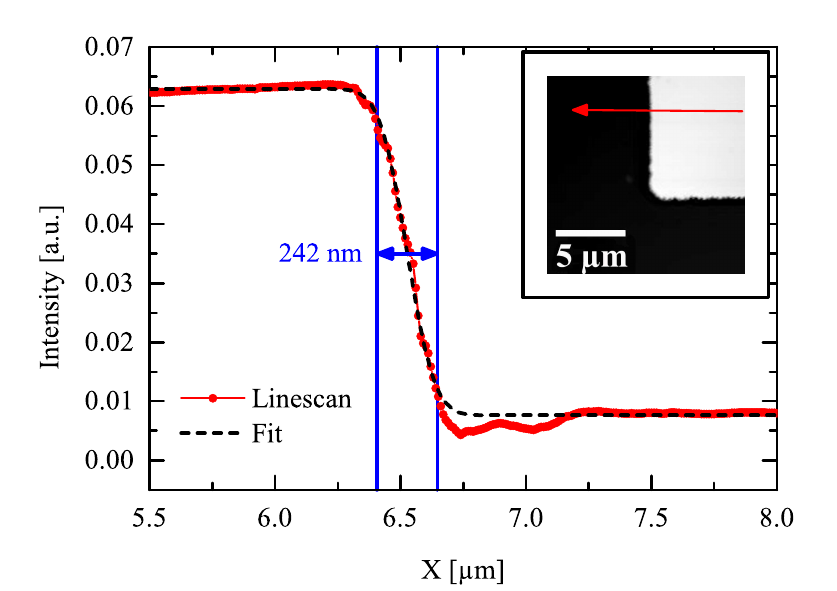}
 \caption{Linescan (red) across the structure shown in the inset. The red arrow indicates the position and direction of the linescan. The data has been fitted (dashed black line) using Eq.~\ref{eq:resfit}. A resolution of $242\,\mathrm{nm}$ is achieved. \label{fig:linescan}}
\end{figure}

A linescan across the edge of a patterned structure is shown in Fig.~\ref{fig:linescan}. A confocal pinhole of $0.5\,\mathrm{AU}$ diameter was used for the acquisition of the image and the linescan.
The intensity profile of the linescan can be used to evaluate the width of the PSF and the corresponding resolution.
For a Gaussian laser beam, the PSF has a Gaussian profile with maximum intensity $I_0$ and $1/\mathrm{e}$-width $\omega$
\begin{equation}
I(x,y)=I_0 \,\mathrm{e}^{-\frac{(x^2+y^2)}{\omega^2}}~.
\end{equation}
The edge-spread function (ESF), obtained by scanning over a sharp edge at $x=0$, is the convolution of the PSF at position $X$ and the edge profile defined by a reflectance $R_1$ for $x<0$ and $R_2$ for $x\geq0$. The ESF is given by 
\begin{align}\label{eq:resfit}
&\begin{aligned}
P(X)=\pi\omega^2I_0R_1&-R_1\int\limits_{-\infty}^0 \int\limits_{-\infty}^{\infty} I(x-X,y)\,dx\,dy \nonumber\\
 &+ R_2\int\limits_0^{\infty} \int\limits_{-\infty}^{\infty} I(x-X,y)\,dx\,dy 
\end{aligned} \\
 &=\pi\omega^2I_0\left[\frac{R_1}{2}\left(1-\mathrm{erf}\frac{X}{\omega}\right)+\frac{R_2}{2}\left(1+\mathrm{erf}\frac{X}{\omega}\right)\right]~.
\end{align}
A resolution criterion for two Gaussian PSF at a distance $d$, that is similar to the Rayleigh criterion, is realized when their combined intensity profile shows a $\approx26\,\%$ dip. This distance is related to the $1/\mathrm{e}$-width of the PSF by $d_{Rayleigh}\approx 1.97\,\omega$.
To determine the resolution, we fitted the linescan in Fig.~\ref{fig:linescan} using Eq.~\ref{eq:resfit} and obtained a value for the $1/\mathrm{e}$-width of the PSF of $\omega=122.6\,\mathrm{nm}$. the resolution, according to the Rayleigh criterion, is found to be $d_{Rayleigh}\approx242\,\mathrm{nm}$, which is close to the theoretical value (Eq.~\ref{eq:cfres}).

\subsection{Detector}
\label{subsec:detector}
Detectors featuring a high sensitivity for the orientation of the plane of polarization $\theta$ are needed for polarized light microscopy. A useful measure to assess the detector sensitivy is the noise spectral density for the orientation of the plane of polarization $S_{\theta}^{1/2}$ , given in $\mathrm{rad}/\sqrt{\mathrm{Hz}}$.
Different approaches towards the measurement of the polarization of light have been employed for scanning polarizing microscopy. The most basic one is the use of a polarizer that is rotated close to $90\,^\circ$ relative to the polarization of the beam. More advanced designs use photoelastic modulators~\cite{Vavassori2000} or Faraday modulators~\cite{Hornauer1990} to modulate the polarization of the light before it is passed through the analyzer. This generates a signal at the second harmonic of the modulation frequency that is proportional to the Kerr rotation and a signal at the modulation frequency that is proportional to the Kerr ellipticity. Cormier~\textit{et al.}~\cite{Cormier2008} measured a polarization noise of $0.3\,\mathrm{mdeg}$ for an integration time of $50\,\mathrm{s}$ using polarization modulation. This corresponds to a sensitivity of $3.7\times10^{-5}\,\mathrm{rad}/\sqrt{\mathrm{Hz}}$. Another approach is to use a differential detector~\cite{Kasiraj86}. Fla\v{j}sman~\textit{et al.} ~\cite{Flajsman2016} report a sensitivity of $5\times10^{-7}\,\mathrm{rad}$ using a differential detector, however they do not provide information on the bandwidth of this measurement. Spielman~\textit{et al.}~\cite{Spielmann1990} developed a detector based on a Sagnac interferometer. A sensitivity of $1\times10^{-7}\,\mathrm{rad}/\sqrt{\mathrm{Hz}}$ using a Sagnac interferometer was demonstrated by Xia~\textit{et al.}~\cite{Xia2006}.

\begin{figure}
 \includegraphics{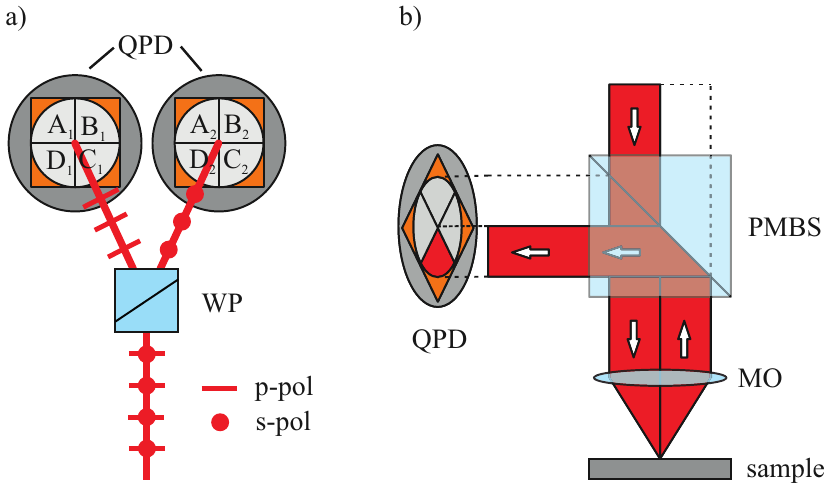}
 \caption{Selection of the illumination path: a) Scheme of the detector: The beam is split into two orthogonally polarized beams by the Wollaston prism (WP). These beams are detected by two quadrant photo diodes (QPD). b) Simplified lightpath for the read out of one quadrant. Each quadrant corresponds to a different illumination direction. \label{fig:detscheme}}
\end{figure}

We decided to use a differential detector based on the design by Clegg~\textit{et al.}~\cite{Clegg1991}. It uses a Wollaston prism to split the incident beam into two beams with orthogonal polarization (Fig.~\ref{fig:detscheme}a) with intensities $I_1\propto I_0 \sin^2\theta$ and $I_2\propto I_0 \cos^2\theta$, for an incident beam with intensity $I_0$ and linear polarization at an angle $\theta$ relative to the s-polarized beam. The orientation of the polarization, the angle $\theta$, can be extracted from the intensities $I_1$ and $I_2$
\begin{equation}
	I_2-I_1 = I_0(\cos^2 \theta-\sin^2 \theta)=I_0 \cos(2\theta)
\end{equation}	
\begin{equation}
	\theta = \frac{1}{2}\arccos \left (\frac{I_2-I_1}{I_0} \right )= \frac{1}{2}\arccos \left (\frac{I_2-I_1}{(I_1+I_2)} \right )~.
\end{equation}	
Note that, with this detection mechanism, the orientation of the plane of polarization can be measured independently from the intensity $I_0$. This differential detection scheme is an important feature, because it cancels to a high degree variations in reflectivity that may occur due to the sample topography. 

The intensity of the two beams is measured using two quadrant photodiodes. Since the microscope objective focusses the beam onto the sample and the quadrant photo diodes are in a conjugate plane to the backfocal plane of the microscope objective, each of the quadrants corresponds to a unique range of angles of incidence , as shown in Fig.\ref{fig:detscheme}b. Furthermore, opposite quadrants correspond to opposite angles of incidence. Because the longitudinal MOKE is an odd function of angle of incidence, while the polar MOKE is an even function of angle of incidence, the polar and longitudinal MOKE can be separated by measuring the Kerr rotation individually for every quadrant~\cite{Clegg1991,Ding2000}. For a quadrant X, the plane of polarization is given by
\begin{equation}
	\theta_X =  \frac{1}{2}\arccos \left (\frac{I_{X2}-I_{X1}}{(I_{X1}+I_{X2})} \right ) ,\, X = \{A,B,C,D\}~.
\end{equation}	
In total, six different signals can be measured simultaneously. The polar Kerr signal $\theta_{K}^{polar}$ which is sensitive to the out-of-plane magnetization
\begin{equation}
	\theta_{K}^{polar}= \theta_A + \theta_B + \theta_C + \theta_D~,
\end{equation}	
the longitudinal Kerr signal $\theta_{K,ij}^{long}$ which is sensitive to the magnetization component along the $ij=\{xy,x\bar{y},x,y\}$ direction in the image plane
\begin{equation}
	\theta_{K,x\bar{y}}^{long} = \theta_A - \theta_C 
\end{equation}
\begin{equation}		
	\theta_{K,xy}^{long} = \theta_D - \theta_B 
\end{equation}
\begin{equation}		
	\theta_{K,y}^{long} = (\theta_C + \theta_D) - (\theta_A + \theta_B)
\end{equation}
\begin{equation}	
	\theta_{K,x}^{long} = (\theta_A + \theta_D) - (\theta_B + \theta_C) 
\end{equation}	
and the reflected intensity (conventional image)
\begin{equation}
	I_{tot}= \sum_{X=\{A,B,C,D\}} I_{X1}+I_{X2}~.
\end{equation}

For reasons of noise suppression and dc error cancellation of the electronics, the measurement is performed in a lock-in configuration where the laser is modulated with a frequency of several kHz and a lock-in amplifier is used to extract the signal. The photocurrent from each quadrant is converted to a voltage using a transimpedance amplifier  with a gain of $G_{TI}=2\,\mathrm{MV/A}$. A second programmable gain amplifier (PGA) stage provides additional gain from $G=1$ to $G=8000$. The PGA is ac-coupled by inserting a high-pass filter with a cut-off frequency $f_c=200$\,Hz at the input of the PGA. The output of the PGAs is recorded with a sampling rate of 2\,MHz and a resolution of 16\,bit by a simultaneous sampling data acquisition card. The signal is demodulated using a software-based lock-in algorithm.
\subsection{Noise Considerations}
\label{subsec:noise}
The different noise sources contributing to the overall system noise can be divided into three main parts: laser intensity noise, detector noise and mechanical noise.
The laser intensity noise consists of the shot noise and excess noise related to the pump current, mode hopping, thermal fluctuations, etc. The excess noise can be reduced by measuring the output power and using it as feedback to control the pump current, as well as stabilizing the temperature of the laser diode. The shot noise, however, cannot be overcome and is a fundamental limit to the intensity noise. Since the orientation of the plane of polarization is measured using a balanced detector, laser intensity noise is cancelled to a high degree.

\begin{figure}
 \includegraphics{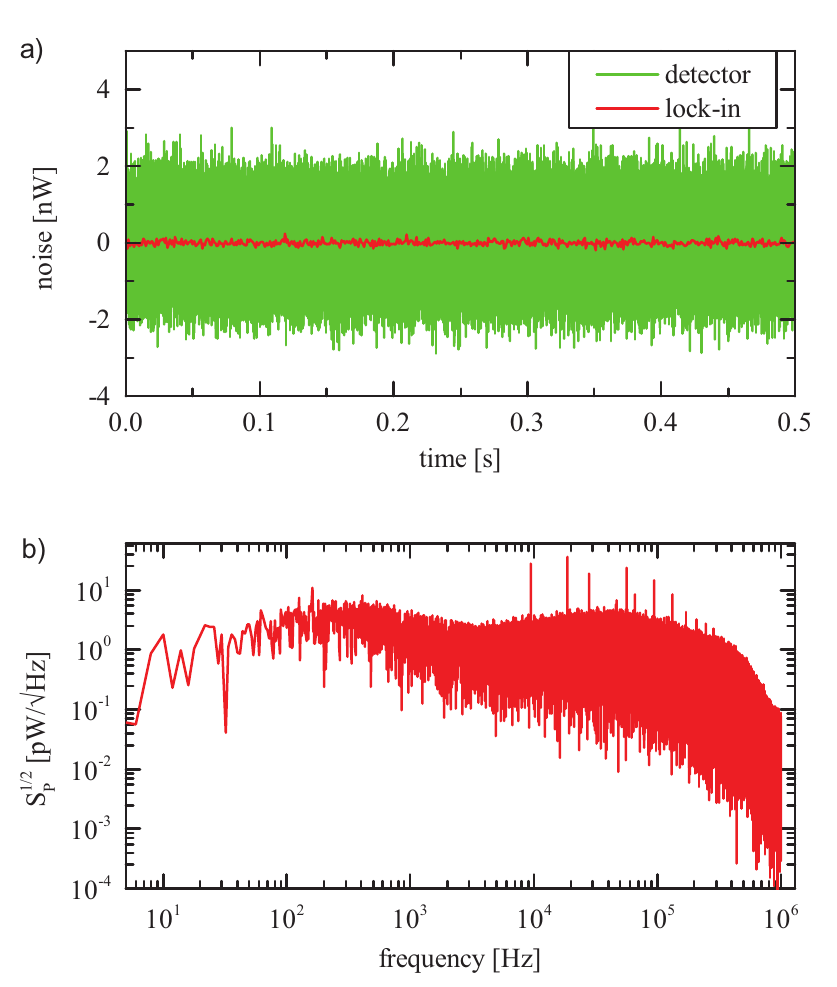}
 \caption{Noise characteristics of the detector: a) detector noise at the PGA output (green) and after lock-in detection with $f_{ref}=10\,\mathrm{kHz}$, $t_{int}=1\,\mathrm{ms}$ (red). Lock-in detection reduces noise by more than an order of magnitude. b) noise spectral density at the PGA output. $1/f$ noise is elliminated by ac-coupling the PGA.\label{fig:detnoise}}
\end{figure}

The detector noise is produced by the electronic components within the detector and does not depend on the signal reaching the detector. The dark noise of one quadrant of the detector, measured at the output of the PGA, is shown by the green curve in Fig.\ref{fig:detnoise}a. The dark noise at the output of the digital lock-in amplifier, for a reference frequency of $f_{ref}=10$\,kHz and a integration time of $t_{int}=1\,\mathrm{ms}$ is shown in red. Using lock-in detection shifts the signal to higher frequencies, where the $1/f$ part of the noise is negligible. The rms noise at the output of the lock-in amplifier is proportional to the detection bandwidth, which is controlled by the lock-in integration time. In the case of frequency independent white noise, it is calculated by dividing the noise spectral density at the PGA output by the square root ot the integration time $\Delta P_{rms}=S_P^{1/2}/ \sqrt{t_{int}}$. The noise spectral density $S_P^{1/2}$ for the PGA output is shown in Fig.\ref{fig:detnoise}b. Noise below the cut-off frequency ($f_c=200$\,Hz) of the ac-coupling filter is suppressed and $1/f$ noise is practically eliminated. The lock-in detection results in a noise reduction of more than one order of magnitude that can be further increased by increasing the integration time $t_{int}$.
 
An additional noise source is mechanical noise, which is related to perturbations to the imaging path caused by vibrations, noise in the angular position of the FSM, vibrations of the coldfinger, etc. We measured a value of $5\times10^{-6}\,\mathrm{rad}/\sqrt{\mathrm{Hz}}$ for the sensitivity to the orientation of the plane of polarization.
\section{Imaging of electric transport properties}
\label{sec:electrictransport}
Imaging of electric transport properties using low-temperature scanning electron microscopy (LTSEM) has first been demonstrated by St\"{o}hr~\textit{et al.}~\cite{Stöhr1979}. In LTSEM, the electron beam generates a local perturbation to the electric transport characteristics of typically a current-biased sample that leads to a global voltage response, which serves as image contrast. A general response theory for this imaging mechanism was developed by Clem~\textit{et al.}~\cite{Clem1980}. For a detailed description of the LTSEM technique see the reviews by Huebener~\cite{Huebener1988} and by Gross~\textit{et al.}~\cite{Gross1994}. A similar technique using a focussed laser beam to perturb the sample was used by Divin~\textit{et al.}~\cite{Divin1991, Divin1994}. It was shown by Dieckmann~\textit{et al.}~\cite{Dieckmann1997} that this technique, low-temperature scanning laser microscopy (LTSLM), delivers results that are equivalent to LTSEM. LTSLM has, among others, been applied to research on superconductors~\cite{Sivakov1996, Werner2013,WangHB2009,Benseman2015}, the quantum Hall effect~\cite{Shashkin1997}, spintronics~\cite{Wagenknecht06, Werner2011} and spin-caloritronics~\cite{Weiler2012}.
\begin{figure}
 \includegraphics{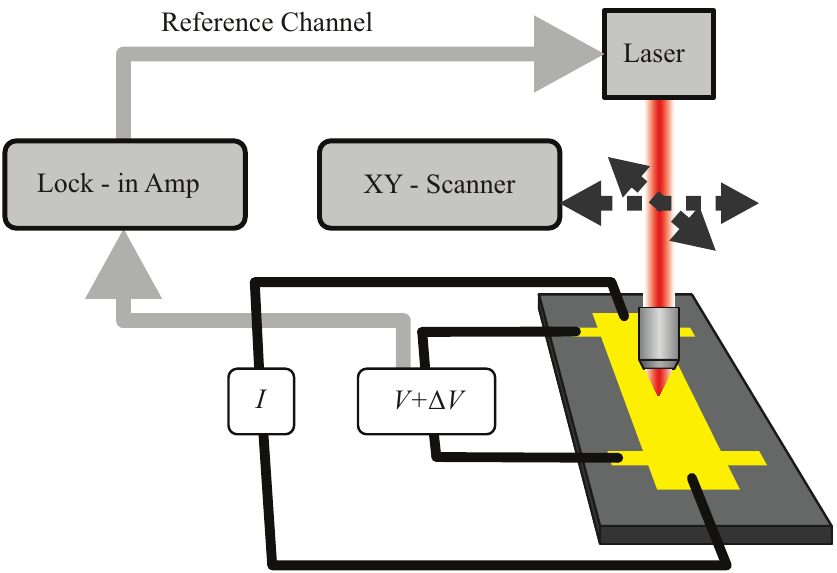}
 \caption{Principle for LTSLM imaging of local electric transport properties. The laser beam, that is intensity-modulated at the reference frequency of the lock-in amplifier, locally perturbs the electric transport properties of the current-biased sample. The perturbation leads to a change $\Delta V$ in global voltage, which is detected by the lock-in amplifier. \label{fig:voltimage}}
\end{figure}
The LTSPM can be operated in LTSLM mode to gain information on the electric transport properties of a sample. The intensity-modulated laser beam generates a periodic perturbation that leads to a periodic global voltage response $\Delta V$, which can be measured with a lock-in amplifier (Fig.~\ref{fig:voltimage}). Maps of the voltage response $\Delta V$, the voltage-image, can be acquired by scanning the laser across the sample. Usually, the primary nature of the perturbation is local heating~\cite{Zhuravel2006b}, although additional mechanism such as the photovoltaic-effect, photoconductivity or quasiparticle creation in superconductors~\cite{Zhuravel2003} are possible and may need to be considered for the interpretation of these images. The spatial resolution of this technique is governed by the length scale on which the laser beam induced perturbation decays. In the case of local heating this is the thermal decay length~\cite{Gross1994} and a typical resolution of a few $\mu$m can be achieved.
\section{Examples}
\label{sec:Examples}
In this Section, we present a few examples demonstrating the capabilities of the system. We will not give a detailed description of the underlying physics for each of the examples as this would go beyond the scope of this paper. However, we include a short motivation and description for each study.    
\subsection{Magnetic domains in Barium Hexaferrite}
\label{subsec:BaFeO}
\begin{figure}
 \includegraphics{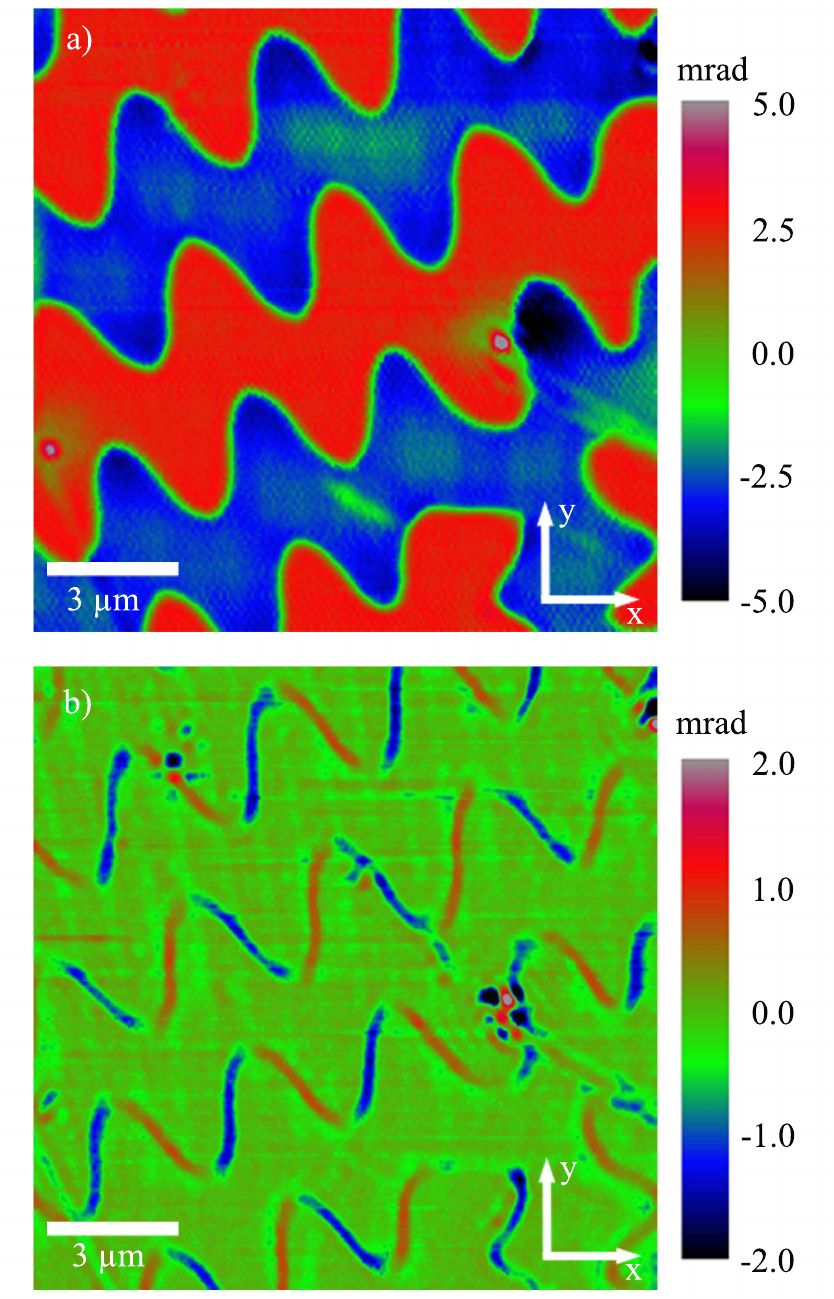}
 \caption{Magnetic domains in BaFe$_{12}$O$_{19}$ observed via the polar MOKE $\theta_{K}^{polar}$ (a) and the longitudinal MOKE $\theta_{K,y}^{long}$ (b) at room-temperature and zero magnetic field.\label{fig:BaFeO}}
\end{figure}

Magnetic domains in the basal plane of a barium hexaferrite (BaFe$_{12}$O$_{19}$(0001)) crystal observed with the LTSPM at room-temperature and zero magnetic field are shown in Fig.~\ref{fig:BaFeO}a) and b). This ferromagnetic material has been extensively studied by magnetic force microscopy and scanning Hall probe microscopy~\cite{Yang2004, Yang2006, Yang2011} as a substrate for superconductor-ferromagnet hybrids. The magnetic domains in BaFe$_{12}$O$_{19}$(0001) form a labyrinth pattern with zigzag domain-walls~\cite{Yang2011} in the remanent state. Due to its uniaxial anisotropy, the magnetization within the domains is parallel/anti-parallel to the [0001]-direction~\cite{Yang2006}. The width of the Bloch-type domain walls was found to be around 200\,nm~\cite{Yang2004, Yang2006}. For a superconductor-ferromagnet hybrid, where a superconducting film is deposited on top of a ferromagnetic substrate, the magnetic landscape within the superconducting film that is generated by the ferromagnetic substrate modifies the nucleation of superconductivity. This can lead to reverse-domain and domain-wall superconductivity, which has been imaged by LTSLM~\cite{Fritzsche06, Werner11a}. 
The polar MOKE ($\theta_{K}^{polar}$) image  is shown in Fig.~\ref{fig:BaFeO}~a). The zigzag folding of the domain walls between domains with magnetization parallel/anti-parallel to the z-direction can clearly be seen. Fig.~\ref{fig:BaFeO}~b) is acquired at the same time and displays the longitudinal MOKE ( $\theta_{K,y}^{long}$) image, which is sensitive to the magnetization component along the y-direction. No contrast is obtained between adjacent domains, where the magnetization points in the out-of-plane/into-the-plane direction. However, a magneto-optical signal is obtained along the Bloch domain-walls, where the magnetization has an in-plane component along the direction of the domain-wall. Since the displayed longitudinal MOKE signal $\theta_{K,y}^{long}$ is only sensitive to the y-component of the magnetization, the contrast is best if the domain-wall runs in y-direction and is lost if the domain-walls run in x-direction.
\subsection{Magnetic flux structures in a superconducting niobium coplanar waveguide resonator}
\label{subsec:Nbresonator}
\begin{figure}
 \includegraphics{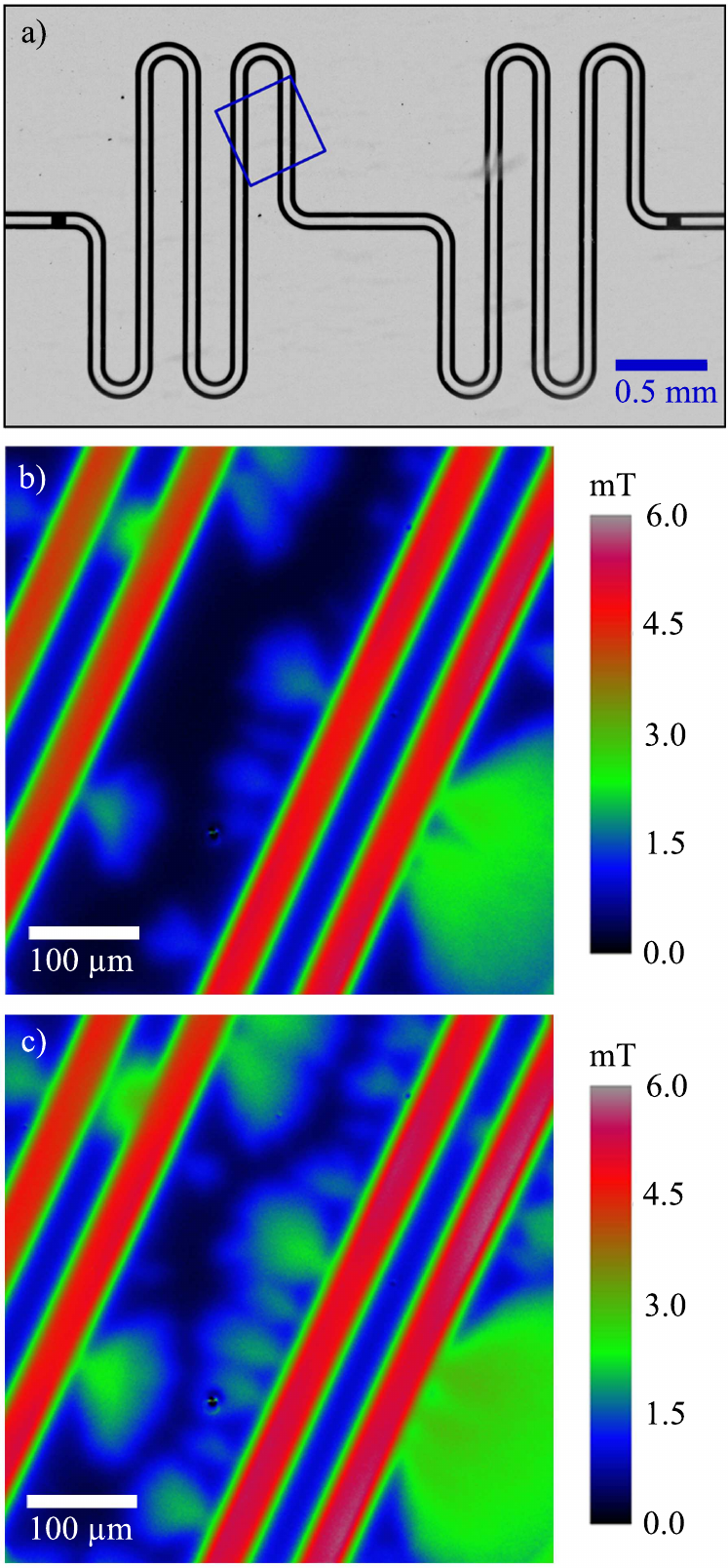}
 \caption{Optical image of a superconducting niobium half-wavelength resonator (a). The ground plane and center conductor appear bright, the gaps between them appear dark. The blue square indicates the area that has been imaged in (b) and (c) using a MOIF and the LTWPM. Magnetic field distribution above the resonator at an externally applied field of $B_{\perp}=1.4$\,mT (b) and $B_{\perp}=1.8$\,mT (c) at a temperature of 5.3\,K.\label{fig:resonator}}
\end{figure}

The performance of superconducting niobium coplanar half-wavelength resonators for hybrid quantum systems in perpendicular magnetic fields is detrimentally affected by the presence of Abrikosov vortices~\cite{Bothner2011, Bothner12a, Bothner12b}. The motion of Abrikosov vortices leads to energy dissipation and hence to increased losses that reduce the quality factor of the resonator. The distribution of magnetic flux within the superconducting resonator can be visualized by placing a magneto-optic indicator film (MOIF)~\cite{Goernert2010, johansen2004} on top of the resonator. The Faraday effect in the MOIF leads to a rotation of the plane of polarization that is proportional to the magnetic field at the position of the MOIF. Fig.~\ref{fig:resonator}~a) shows an optical image of a capacitatively coupled niobium half-wavelength resonator. The ground planes and the center conductor are the bright parts, while the gaps between them appear dark. We investigated the penetration of magnetic flux into the part of the resonator that is highlighted by the blue square using the LTWPM and a MOIF. The MOIF used for this study consists of a $4.9\,\mu$m thick bismuth substituted rare-earth (RE) iron garnet, of composition (Bi,RE)$_3$(Fe,Ga)$_5$O$_{12}$, grown by liquid phase epitaxy on a gadolinium gallium garnet (Gd$_3$Ga$_5$O$_{12}$) substrate. The MOIF is covered with a mirror layer and a $4\,\mu$m thick protective layer of diamond-like carbon. Because of this relatively thick protective layer, the MOIF is at a distance to the superconductor where it is not possible to resolve the stray field of individual vortices. We expect that we will reach single vortex resolution with improved MOIF and that the LTSPM can be used to manipulate vortices, as has been demonstrated by Veshchunov~\textit{et al.}~\cite{Veshchunov2016}. 

To obtain calibration data, the magneto-optic response of the MOIF was measured at $10\,\mathrm{K}$, where the niobium film is in the normal state. The resonator was cooled to a temperature of 5.3\,K in zero magnetic field and a reference image was aquired, before the magnetic field was increased. For the subsequent acquisitions, it is possible to convert the raw image data to magnetic field maps, by subtracting the reference image and rescaling the pixel values according to the calibration data that has been acquired at $10\,\mathrm{K}$. This was done using the image processing software FIJI~\cite{Schindelin2012}. Fig.~\ref{fig:resonator}~b) shows the magnetic field distribution at an externally applied magnetic field, with an orientation perpendicular to the substrate surface, of $B_\perp=1.4$\,mT. The magnetic field is focussed through the gap between the center conductor and the ground planes and magnetic flux in the form of Abrikosov vortices has entered the superconducting film. A few locations along the edge of the niobium film are favored for flux entry, indicating a locally reduced surface barrier for flux penetration. The vortices that enter at these sites tend to get pinned at defects in the superconducting film. This results in a highly non-homogeneous distribution of Abrikosov vortices in the niobium film. As the magnetic field is increased to $B_\perp=1.8$\,mT (Fig.~\ref{fig:resonator}~c)) further flux entry sites appear and additional vortices enter the superconducting film. A comprehensive study of magnetic flux penetration in niobium thin films and the effect of indentations and roughness at the film border has been carried out by Brisbois~\textit{et al.}~\cite{Brisbois2016}.
\subsection{Twin-walls between ferroelastic domains in SrTiO$_3$ and their effect on electric transport at the SrTiO$_3$/LaAlO$_3$ interface}
\label{subsec:LAOSTO}
The interface between LaAlO$_3$ (LAO) and SrTiO$_3$ (STO) has attracted considerable interest over the past years. Although both materials are insulating in bulk, the interface between a LAO-layer of at least four unit cells thickness and its TiO$_2$-terminated STO-substrate becomes electrically conducting~\cite{Ohtomo04}. Upon cool-down, STO undergoes a ferroelastic phase transition from cubic to tetragonal at around 105\,K. The presence of twin-walls between ferroelastic domains in tetragonal STO has an influence on the two-dimensional electron gas at the LAO/STO interface, that has been studied by scanning SQUID microscopy~\cite{Kalisky13}, scanning single-electron transistor microscopy~\cite{Honig13} and LTSEM~\cite{Ma2016}. Imaging of twin-walls in ferroelastic STO is also possible by polarized light microscopy~\cite{Erlich2015}. Birefringence in the STO substrate~\cite{Geday2004} leads to a contrast at the twin-walls between adjacent ferroelastic domains.

\begin{figure}
 \includegraphics{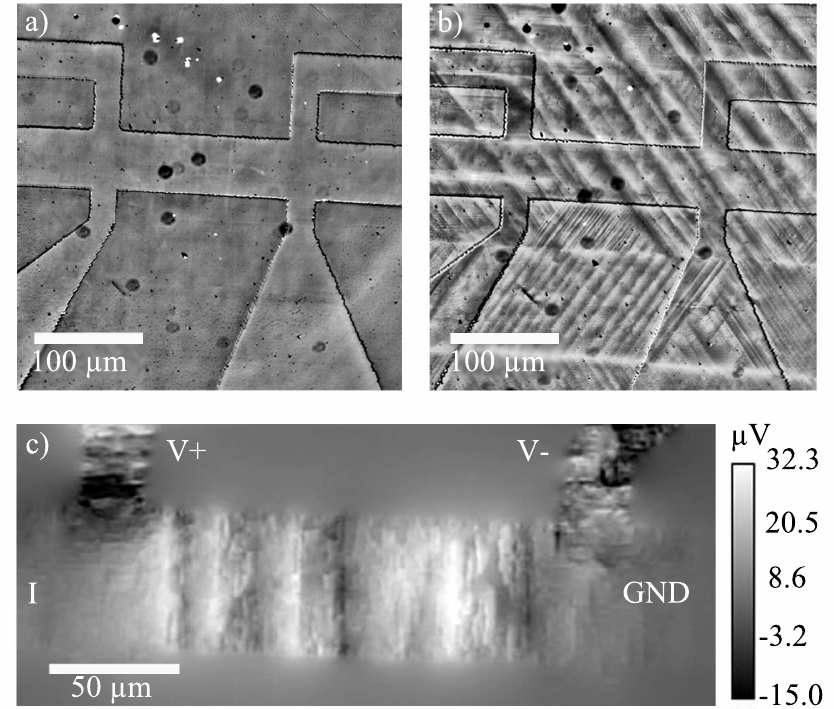}
 \caption{LTWPM images of a LAO/STO(110) Hall bar at (a) 107\,K and (b) 75\,K. Twin walls between ferroelastic domains appear below 105\,K. (c) LTSLM $\Delta V$ image of a Nb/LAO/STO(001) Hall bar at a bias current of 11\,$\mu$A and a temperature of 5\,K.\label{fig:LAOSTO}}
\end{figure}

A LAO Hall bar on a (110)-oriented STO substrate, imaged with the LTWPM at a temperature of 107\,K, is shown in Fig.~\ref{fig:LAOSTO}~a). The long side of the Hall bar is aligned with the [001]-axis of the STO substrate. At 107\,K, the STO is in its cubic phase and no ferroelastic domains can be observed. After cooling the sample through the cubic to tetragonal phase transition of STO, twin-walls between ferroelastic domains in the STO substrate appear, as can be seen for a temperature of 75\,K in Fig.~\ref{fig:LAOSTO}~b). The twin-walls are oriented at angles of $0\,^{\circ}$, $55\,^{\circ}$ and $125\,^{\circ}$ with the [001]-axis of the STO substrate, which corresponds exactly to the expected values calculated in Ref.~\onlinecite{Ma2016}. The images have been corrected for uneven illumination by using the 'subtract background' feature in FIJI~\cite{Schindelin2012}.
A LTSLM $\Delta V$ image of a Nb/LAO/STO(001) sample at a temperature of 5\,K is shown in Fig.~\ref{fig:LAOSTO}~c). The LAO-layer is patterned into a similar structure as in (a) and (b), however the substrate has a different orientation. Since LAO and STO are highly transparent for visible light, a niobium layer has been added on top of the sample to absorb the laser beam, resulting in a local perturbation to the sample. A bias current of 11\,$\mu$A was sent through the two-dimensional electron gas (from I to GND). The voltage response $\Delta V$, due to the perturbation of the laser beam, was measured between V+ and V- using a lock-in amplifier at a reference frequency of 10\,kHz and with a time constant of 1\,ms. Twin-walls in the STO-substrate lead to a modification of the LTSLM response and are visible as stripe-like structures at an angle of $90\,^{\circ}$ with the long side of the Hall bar in the LTSLM image. For this substrate, the expected angles of the twin-walls with the long side of the Hall bar are $0\,^{\circ}$, $45\,^{\circ}$, $90\,^{\circ}$ and $135\,^{\circ}$. Similar investigations have been conducted using low-temperature scanning electron microscopy~\cite{Ma2016}.
%.
%
%
\section{Summary}
\label{sec:summary}
\begin{table}
\caption{system specifications\label{tab:specs} }
 \begin{tabular}{|l|l|l|}
		\hline
		& LTSPM & LTWPM\\ \hline
		resolution & 242\,nm & 481\,nm \\ \hline
		sensitivity & $5\times10^{-6}\,\mathrm{rad/\sqrt{Hz}}$ & $1.0\times10^{-4}\,\mathrm{rad/\sqrt{Hz}}$ \\ \hline
		field of view & \multicolumn{2}{c|}{$500 \,\mathrm{\mu m}\times 500 \,\mathrm{\mu m}$} \\ \hline
		temperature range & \multicolumn{2}{c|}{4\,K to 300\,K} \\ \hline
		$B_\parallel$ & \multicolumn{2}{c|}{ $\pm800$\,mT} \\ \hline
		$B_\perp$ & \multicolumn{2}{c|}{ $\pm20$\,mT} \\ \hline
 \end{tabular}
\end{table}
We have presented a versatile polarizing microscope that offers the possibility to image ferromagnetic, ferroelectric and ferroelastic domains by using either confocal laser scanning (LTSPM) or widefield microscopy (LTWPM). Both imaging modes achieve excellent lateral resolution over a wide field of view. The lateral resolution of the LTSPM of 242\,nm is close to the resolution limit for imaging with visible light. The instrument is equipped with highly sensitive polarized light detectors that provide a sensitivity of $1.0\times10^{-4}\,\mathrm{rad/\sqrt{Hz}}$ for the LTWPM and $5\times10^{-6}\,\mathrm{rad/\sqrt{Hz}}$ for the LTSPM. A $^4$He continuous flow cryostat enables observations at sample temperatures ranging from 4\,K to 300\,K, and magnetic fields with variable orientation can be applied to the sample. The system specifications are summarized in table~\ref{tab:specs}.
We have demonstrated the capability of the microscope to image ferromagnetic domains and domain-walls in BaFe$_{12}$O$_{19}$, ferroelastic domains in SrTiO$_3$, the magnetic field distribution above a superconducting Nb film and electrical transport characteristics of the 2-dimensional electron gas at the interface of LaAlO$_3$ and SrTiO$_3$. 
We expect that the instrument will prove to be useful for investigations of a wide variety of solid-state effects. The combination of magnetic, structural and electric imaging with high lateral resolution and variable temperature has the potential to deliver important insights for research in spintronics, spin caloritronics, superconductivity, magnetism and their hybrid-systems.
\begin{acknowledgments}
% Put your acknowledgments here.
We thank J. Fritzsche for providing the BaFe$_{12}$O$_{19}$ crystal, D. Bothner and B. Ferdinand for providing the Nb resonator, M. Lindner (Innovent e.V. Jena) for providing the MOIF, and H.J.H Ma and A. St\"{o}hr for providing the LAO/STO samples. 
This work was funded by the Deutsche Forschungsgemeinschaft (DFG) via project No. KO 1303/8-1.
\end{acknowledgments}
\bibliography{LTSPMbib_arxiv}
% Run this once to generate your BBL file. Then copy the contents of your BBL file into your main latex file, commenting out "\bibliography"
%
%
%
\end{document}